\def\cm{cm$^{-1}$}
\RequirePackage{fix-cm}
\documentclass[smallextended]{svjour3}       % onecolumn (second format)
\smartqed  % flush right qed marks, e.g. at end of proof
\usepackage{graphicx}
\begin{document}

\title{Molecular Dynamics at Electrical- and Optical-Driven Phase Transitions}
\subtitle{Time-Resolved Infrared Studies Using Fourier-Transform Spectrometers}


\author{Tobias Peterseim \and Martin Dressel}


\institute{Tobias Peterseim and Martin Dressel\at
              1. Physikalisches Institut, Universit\"at Stuttgart,\\
              Pfaffenwaldring 57, D-70550 Stuttgart, Germany \\
              Tel.: +49-711-685 64946\\
              Fax:  +49-711-685 64886\\
              \email{dressel@pi1.physik.uni-stuttgart.de}           %  \\
}

\date{Received: date / Accepted: date}

\maketitle

\begin{abstract}
The time-dependent optical properties of molecular systems are investigated by step-scan Fourier-transform spectroscopy in order to explore the dynamics at phase transitions
and molecular orientation in the milli- and microsecond range.
The electrical switching of liquid crystals traced by vibrational spectroscopy reveals a rotation of the molecules with a relaxation time of 2~ms. The photo-induced neutral-ionic transition in TTF-CA takes place by a suppression of the dimerization in the ionic phase
and creation of neutral domains. The time-dependent infrared spectra depend on temperature and laser pulse intensity; the relaxation of the spectra follows a stretched-exponential decay with relaxation times in the microsecond range strongly dependent on temperature and laser intensity.
We present all details of the experimental setups and  thoroughly discuss the technical challenges.

\keywords{Fourier-transform infrared spectroscopy \and step-scan technique \and
time dependent phenomena \and vibrational spectroscopy \and liquid crystals
\and photo-induced phase transition }
\end{abstract}

\section{Introduction}
\label{intro}
Fourier-transform infrared (FTIR) spectroscopy is a widely utilized method to investigate the optical response of gasses, liquids and solids \cite{Bell72,Gremlich03,Griffiths07,Smith11}. In general, steady-state properties are measured, however, numerous approaches have been developed over the years to explore time-dependent phenomena by Fourier-transform interferometry  \cite{Sloan89,Smith02}, many of them optimized for a certain time regime.
Standard rapid-scan techniques are limited by the mirror velocity to a fraction of a second (typically 10~ms), depending on the spectral resolution $\Delta \nu$ required: $\Delta t \propto 1/\Delta \nu$. Since this is often not sufficient,
step-scanning is nowadays implemented in several high-end commercial Fourier-transform instruments
because there is no inherent time limit. It covers the largest dynamical range
with a time resolution of the order of nanoseconds determined by the current detector and electronics technology \cite{Smith02,Laubereau85,Griffiths99} and still achieving a high spectral resolution $\Delta \nu > 1$~\cm. The only restriction is the repeatability of the process: depending on the resolution, spectral range and desired signal-to-noise ratio, it has to be executed hundreds of times.

In contrast to the continuously moving interferometer mirror in the rapid-scan configuration,
the advantage of the step-scan method is the continuous recording of the signal at a fixed mirror position. This is repeated after the mirror has moved to the next position, eventually
composing the complete interferogram.
The step-scan technique is mainly applied in biophysics and polymer chemistry where, for instance, the photolysis processes of chemical reactions \cite{Johnson93}, bacteria systems \cite{Uhmann91,Hessling97,Rammelsberg97,Rammelsberg99,Schleeger09} and the time-dependent reorientation of liquid crystals under the influence of a short electric field \cite{Gregoriou91,Katayama95,Katayama97,Verma01,Huang06} are studied. But also temperature- and light-induced phase transitions can be investigated \cite{Peterseim15,Peterseim16}. Furthermore, it is used to examine the characteristics of lasers and their mode spectra \cite{Johnson93} and for photo-reflection measurements of semiconducting materials as well as quantum wells \cite{Shao06,Shao07}.

Here we want to present and discuss our experimental setups employed to investigate the dynamics of phase transitions. Tracing the vibrational spectra of molecules, we probe the configurational changes after the transition has been triggered either by a short laser pulse (Sec.~\ref{sec:OpticalSwitching}) or by an electrical pulse (Sec.~\ref{sec:ElectricalSwitching}).

\section{Trigger and Data Acquisition}
\label{sec:Trigger}

\begin{figure}
	\centering
		\includegraphics[width=0.47\textwidth]{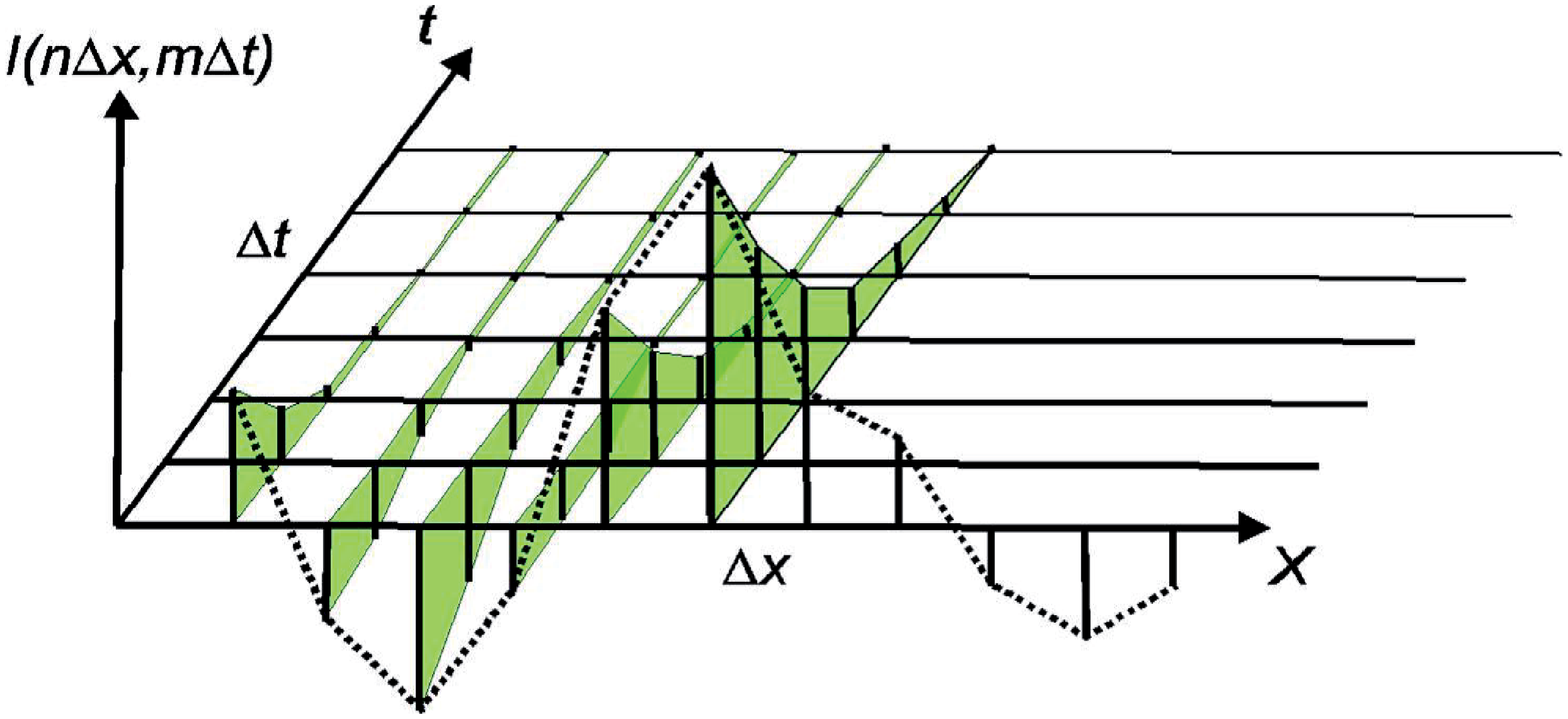}\hspace*{5mm}
		\includegraphics[width=0.47\textwidth]{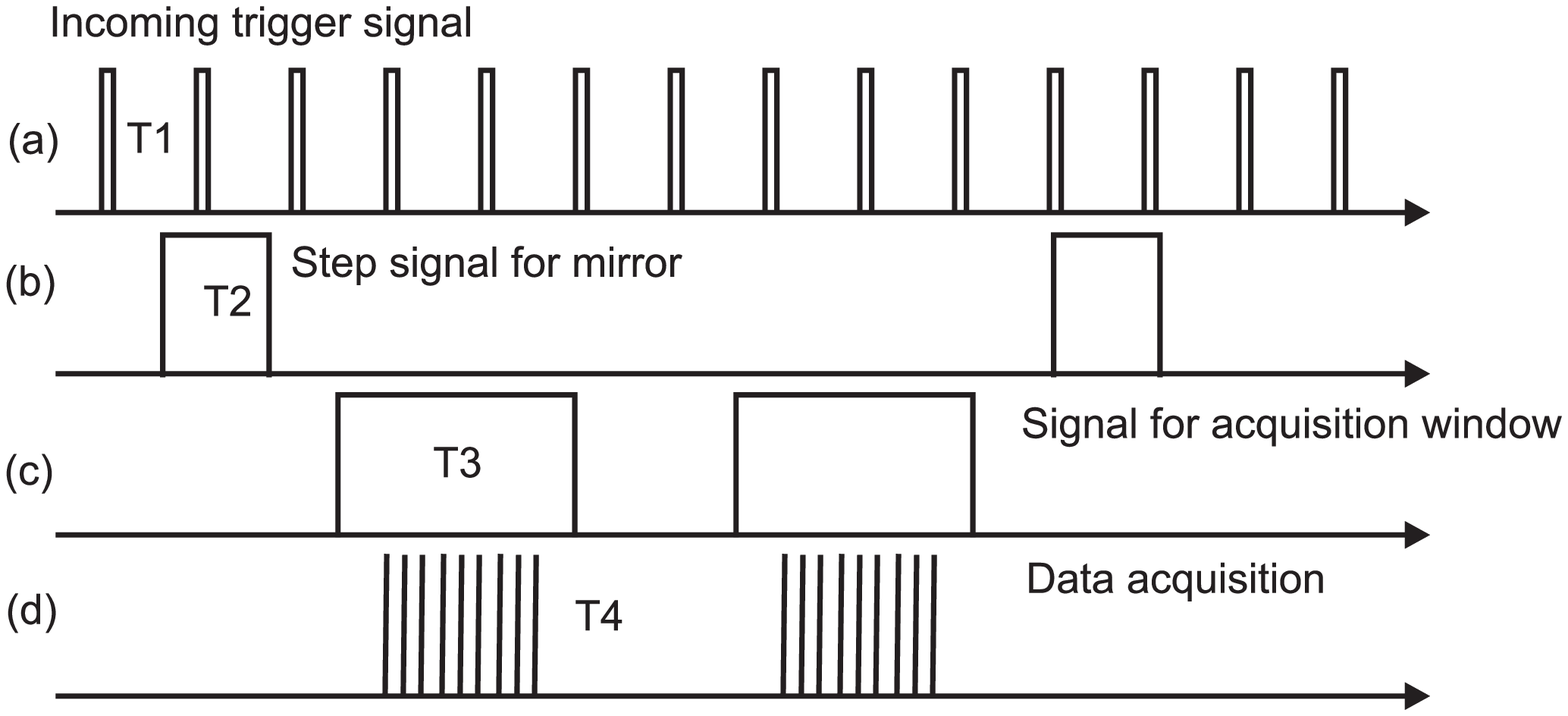}
	\caption{(left panel) At each mirror position $n \Delta x$ the detector signal $I^{\ast}(n  \Delta x,m   \Delta t)$ is recorded as a function of time. After all mirror positions were passed through, the time-dependent signal is recieved by the execution of a Fourier transformation for each measured time point $m \Delta t$.
(right panel)~Illustration of the signal sequence within a step-scan experiment. (a)~External trigger signal T1 given by the external perturbation source, for example a pulse generator or laser. (b)~The second signal T2 is generated by the spectrometer and sent to the interferometer motor to move the mirror to the next position. (c)~After a certain predefined stabilization time the third signal T3 waits for the next external trigger signal T1 and rises afterwards immediately. It stays high as long as all data points are captured. (d)~Signal T4 corresponds to each recorded time point within the signal T3.}
	\label{fig:1-StepScan}	
    \label{fig:2-Trigger}
\end{figure}

In the left panel of Fig.~\ref{fig:1-StepScan} the data acquisition of a step-scan measurement is schematically depicted. At each mirror position $n \Delta x$ along the traveling distance of the mirror the temporal varying reflection signal $I^{\ast}(n \Delta x,m\Delta t)$ is recorded. The complete interferogram is sampled for various times and retardations by successive stepwise moving of the mirror. A subsequent Fourier transformation for each measured time point $m \Delta t$ in the intensity ``matrix'' $I^{\ast}(n  x,m \Delta t)$ derives the time-resolved spectrum $S^{\ast}(j \Delta \nu, m \Delta  t)$.
The process can be repeated several times to improve the signal-to-noise ratio whereas at each $n  \Delta x$ position the time-dependent signal is averaged (typically 10 to 50 times). Additionally, the stability of the mirror influences the signal-to-noise ratio significantly \cite{Uhmann91,Rodig99}, therefore, one must take care that the spectrometer is located in a vibration-free and silent environment. For this reason we place our FTIR-spectrometer on a heavy optical bench mounted on air attenuators to decouple the system from the environment. Furthermore, the vacuum pumps were placed in a separated room to reduce vibrations and acoustic noise. This way we reached a mirror stability better than 3~nm.

Since the data acquisition at each mirror position $n \Delta x$ has to start always at the same time, the data recording must be synchronized with the external stimulation source (laser or pulse generator). Thus, the trigger signal sequence is very important and crucial in a step-scan measurement.
The temporal sequence of the  {TTL} control signals is visualized in the right panel of Fig.~\ref{fig:2-Trigger}. An external or an internal trigger signal T1, which is correlated with the beginning of the reaction, controls the data acquisition. Yet, before the recording starts, the spectrometer sends a signal T2 to the interferometer, so that the mirror is moved to the next position. There, it is stabilized for a few milliseconds.
After the stabilization procedure ($\sim 20$~ms) the next arising trigger signal T1 is used as a starting signal for the record window T3. The signal T3 stays high for the total recording time $T=N \cdot \Delta t$ ($N$ total number of time slices) which is defined at the beginning of each measurement. As soon as the signal T3 is on, at each  {TTL}-signal T4 the detector signal is captured. Depending on the number of averaging spectra this procedure is repeated several times starting again with the T3 signal. Subsequently, the mirror moves to the next position.

The time resolution $\Delta t$ depends mainly on the response time of the detector. Standard PC-MCT-detectors, operating in the photo-current mode, have a minimum response time of $1~\mu$s. Their disadvantage is that the measured current becomes nonlinear above a certain incident threshold intensity. A photovoltaic (PV) detector has a two order of magnitudes shorter response time due to the small detector area.
%it stays linear due to the low applied bias voltage of 100~mV.
Furthermore, the measured signal is always proportional to the incident light intensity. For the measurement a PC-MCT of the model D316 (Bruker Optics, Ettlingen) with a time resolution of about 1~$\mu$s and a PV-MCT KMPV11-11-1-J1 from Kolmar Technologies with a theoretical rise time of 25~ns are utilized. The time resolution also depends on the amplifier and the A/D-converter. There are two amplifiers, the build-in amplifier of the spectrometer and the KA100-A1 from Kolmar Technologies with a bandwidth of 250~MHz. As an A/D-converter the internal converter of the spectrometer with its time resolution of 6~$\mu$s, with a dynamical range of 24~bit and a maximal input voltage of $V_{\rm pp}= 20$~V, can be used, or a transient recorder M3i.4142 from Spectrum Systementwicklung Microelectronic  GmbH, Grosshansdorf with a bandwidth of 400~MHz at 16~bit and $V_{\rm pp}= 10$~V.

The standard interferogram of a Fourier-transform spectrometer consists of an ac- and dc-component: while only the ac-signal contains the important spectral information the dc-signal is usually removed by an electronic high-pass filter. The phase correction can only be performed directly for a rapid-scan measurement. In a step-scan experiment the spectrum can include positive as well as negative features. Thus, a phase correction with the raw time-resolved ac-signal does not work. Two options exist to resolve this complication: first, the simultaneously recording of the dc-signal and the ac-signal; the dc-component provides the right phase correction for the ac-component. The second possibility is to use the phase of a previous rapid-scan measurement \cite{Griffiths07,Herres84,Peterseim15d}.

\section{Electrical Switching}
\label{sec:ElectricalSwitching}
Electrically induced phenomena occur mainly in semiconductor leading to non-linear conductivity or negative differential resistance \cite{Sze07,Yu10}. High electric fields can also trigger phase transitions, for instance in charge-density wave systems \cite{Gruner94} or correlated insulators \cite{Asamitsu97,Sawa08}. In organic conductors electrical switching is subject of research for quite some time \cite{Potember79,Tokura88,Kumai99,Inagak04,Mori07,Ivek12,Peterseim16}.

As an example of electric switching, we here have investigated the polari\-za\-tion-dependent vibrational modes
of liquid crystals, which allow us to trace the orientation of the molecule upon applying a voltage by
time-resolved infrared measurements.
In these molecules several vibrational modes exist which can be assigned to different parts of the molecules. By static and time-resolved polarization-dependent measurements we gain on the one hand information about the orientation of the individual molecular building blocks in static positions and on the other hand about the temporal evolution of the electric switching process. Hence, we learn about the rotation of the different molecular constituents with the electric field.

\subsection{Liquid Crystals}
\label{sec:LiquidCrystals}

\begin{figure}
	\centering
		\includegraphics[width=0.8\textwidth]{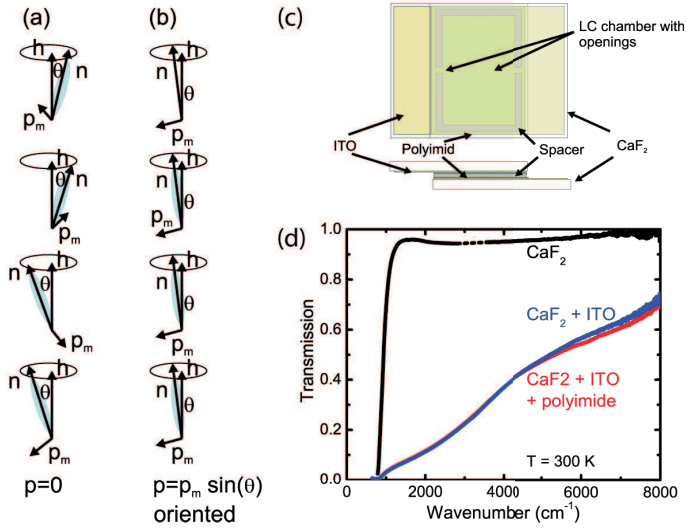}
	\caption{(a)~Chiral smectic phase (SmC$^*$): going from layer to layer the director $\vec{n}$ successively helically precedes around the stacking normal vector $\vec{h}$ in an angle $\theta$. The spontaneous polarization $\vec{p}_m$ is perpendicular to the director and the normal vector. The total polarization $\vec{p}_{\rm total}$ is zero. (b)~Oriented smectic C phase: all directors $\vec{n}$ are aligned towards the same direction. By symmetry the polarization  $\vec{p}_m = p_m \cdot (\vec{h} \times \vec{n})=p_m \sin{\theta}$ can adopt two possible values. (c)~Side- and top view of the liquid crystal cell consisting of two CaF$_2$ plates (light gray, transparent). A thin film of ITO was sputtered onto the plates (bright brown). Polyimid (green) was deposited by the spin-coating process on top of the ITO layer and serves as an orientation layer for the liquid crystals. UV-glue in connection with small plastic spheres (shaded area) with a thickness of $d=5\,\mu$m keeps the two windows at a fixed distance. (d)~Transmission spectrum of CaF$_2$(black) measured at room temperature in comparison to CaF$_2$ coated with ITO (blue) and CaF$_2$ in combination with ITO and polyimid (red). The dashed line indicates a spectral range where water absorption and ice on the detector window corrupt the data.}
	\label{fig:3-LC}	
\end{figure}

Liquid crystals are widely used for electronic displays due to the possibility to manipulate the orientation of the complex organic molecules by applying an electric field and this way to control the transmitted light.
We have chosen the FELIX 017/100 mixture from the Clariant as a ferroelectric liquid crystal that constitutes a chiral smectic C phase (SmC$^{\ast}$) at ambient conditions creating a helical structure with $\theta=14^{\circ}$, as depicted in Fig.~\ref{fig:3-LC}(a). The spontaneous electric polarization $\vec{p}_m=p_m \cdot (\vec{h} \times \vec{n})=p_m \sin{\theta}$ is perpendicular to the director vector $\vec{n}$  and the layer normal vector $\vec{h}$. While in general the different orientations average and the total polarization is zero, a uniaxial rubbed surface aligns the molecules parallel to each other, leading to a net polarization, sketched in Fig.~\ref{fig:3-LC}(b). The whole procedure is well-known as ``surface stabilized ferroelectric liquid crystal'' (SSFLC) \cite{Clark80}.
Two energetically equal configurations exist with $p_m$ arranged in opposite directions, but perpendicular to the surface with the same tilting angle $\pm\theta$. An applied electric field can now induce the transition between the two polarization states by changing the orientation of the molecule.

%\paragraph{Cell}
While standard liquid-crystal cells consist of glass windows transparent in the visible, we have constructed a cell with two CaF$_2$ windows suitable for the infrared spectral range from 1000~\cm\ to 50\,000~\cm\  \cite{Palik09}. In Fig.~\ref{fig:3-LC}(d) the transmission spectrum of a CaF$_2$ window is plotted in the mid-infrared frequency range.
Compared to alternative window materials, such as KBr, it is not hydroscopic, and with $n=1.4$ has a much lower refractive index than Ge ($n=4$), for instance, reducing the reflection losses considerably.
The CaF$_2$ windows ($20\times 20$~mm$^2$,  thickness of $0.1$~mm, purchased from
Korth Kristalle GmbH, Kiel) are polished optically to get the optimal surface quality for the optical experiments.

Both CaF$_2$ windows are sputtered with a thin film of indium tin oxide (ITO) commonly used as an electrode material for displays because it is lucent over a broad energy range and at the same time conductive. However, optical measurements plotted in Fig.~\ref{fig:3-LC}(d) reveal that the transmission decreases rapidly towards low frequencies \cite{Biswas03}. In the range of interesting for the C=C double bond vibrations the transmission is between 5 and 20\%, which is still sufficient for our experiment.
In order to preorient the liquid-crystal molecules a very thin film of polyimide was deposited on top of the ITO layer by using the spin-coating technique which was rubbed afterwards unidirectional [Fig.~\ref{fig:3-LC}(c)]. As soon as the molecules were placed on the surface, they align themselves along the rubbing direction \cite{Sakamoto94}. As demonstrated in Fig.~\ref{fig:3-LC}(d) the very thin layer of polyimide does not affect the overall light transmission of the system in the relevant frequency range in agreement with previous results \cite{Gregoriou91,Verma97,Huang06}.
Both plates have been clued together with a UV glue mixed with small macroscopic plastic spheres with a diameter of $d=5\,\mu$m which serve as spacer to keep the windows on a constant distance, as sketched in Fig.~\ref{fig:3-LC}(c). After the glue was cured, the cell was heated up above $80^{\circ}$C so that the liquid-crystal mixture transformed in the isotropic phase with a reduced viscosity. The liquid crystals were sucked in the chamber due to the capillary action. Afterwards, the temperature of the melt was slowly lowered so that a single domain was formed. Due to the large dimension of the cell the distance of the two plates was not constant and reduces to the center of the cell causing Newton's rings. It changes again, when placed in a vacumm spectrometer. The contact wires were made out of copper and fixed on the exposed ITO side areas with indium; they are connected to a Philips-PM 5768B pulse generator to switch the liquid crystals. The generator sends during the measurement simultaneously to the switching voltage pulse a second TTL signal to the spectrometer to start the data acquiring process.

%\paragraph{Switching of Molecules}
The cell is placed in a Bruker Vertex 66v/s Fourier-transform infrared spectrometer together with a suitable infrared polarizer. The absorption becomes maximal if the infrared light is polarized parallel to the electronic transition dipole moment $\mu_e$ of the  molecular vibrational mode. As soon as the external electric field $E$ is switched on, a force acts on the liquid crystal. In general, the switching process takes a few microseconds $\tau_{\rm or}$ until most of the molecules are reoriented. The new orientation of the molecules and its director $\vec{n}$ change the direction of the transition dipole moment $\mu_e$ leading to an increase or decrease of the absorption signal.

If the electric field is switched on, the directors shown in
Fig.~\ref{fig:3-LC}(b) spin collectively in one direction and therefore, the azimuthal angle changes. The switching velocity and -time can be derived from the equation of motion \cite{Clark83,Handschy88,Goodby14}:
\begin{equation}
\eta \dot{\chi}(t,z)=-I \ddot{\chi}(t,z)+K\nabla^2 \chi(t,z) -p E \sin{\chi(t,z)} \quad,
\label{eq:Eq_LCD1}
\end{equation}
with $\eta$ the damping due the viscosity of the liquid,
$\chi$ is the angle between the electric field and the polarization of the molecule, $z$ is the coordinate along the direction perpendicular to the window, and $K$ the elasticity constant. Due to the large moment of inertia $I_m=1\times 10^{-16}$~kg/m the equation can be simplified to $\eta \dot{\chi}=K\nabla^2 \chi -p E \sin{\chi}$ and solved for small angles $\chi$
\begin{equation}
\chi=\chi_0 \exp\left\{\frac{t}{\tau}\right\}\cdot \sin\left\{\frac{\pi z}{d}\right\}
\end{equation}
with $d=5~\mu$m as the spacing between the two cell windows. We can extract two time constants, $\tau_{\rm or}$ for rise with the field and $\tau_{\rm reor}$ for fall after the electric field $E$ is turned off:
\begin{equation}
\tau_{\rm or}=\frac{\eta}{pE-K\frac{\pi^2}{d^2}} \hspace*{2cm}
\tau_{\rm reor}=\frac{\eta d^2}{K\pi^2}
\label{eq:tau} \quad .
\end{equation}
The switching time for the here examined material FELIX 017/100 can be estimated using
$\eta=8 \times 10^{-6}$~Ns/cm$^{2}$, $p_m=47$~nC/cm$^{2}$, $K=5\times 10^{-12}$~N, $E=10$~kV/cm, and $d=5~\mu$m to yield $\tau_{\rm or}=170~\mu$s and $\tau_{\rm reor}=40$~ms.

\subsection{Steady-State Optical Properties}
\begin{figure}[b]
	\centering
		\includegraphics[width=0.7\textwidth]{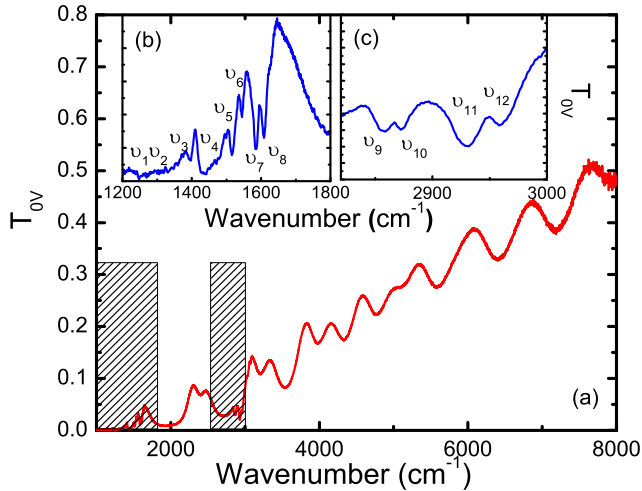}
	\caption{(a)~Transmission spectrum of FELIX 017/100 measured at room temperature. The oscillations originate from interference effects caused by multi-reflections within the liquid crystal cell and windows. (b)~The detail view of the frequency region which shows mainly resonance features of the vibrations of the molecular body. (c)~The inset shows the feature of the methyl vibrations.}
	\label{fig:4-Transmission}	
\end{figure}

The mid-infrared transmission spectrum of the liquid crystal cell filled with FELIX is displayed Fig.~\ref{fig:4-Transmission}.
The dips in the spectrum  between 1000~\cm\ and 1700~\cm\
as well as between  2800 and 3000~\cm, enlarged in
the insets (b) and(c), be assigned to the vibrational modes of the liquid-crystal molecules. The features $\nu_{9}-\nu_{12}$ are stretching modes of CH$_2$- and CH$_3$ end group.  The modes $\nu_4 - \nu_8$ are connected with the vibrations of the C=C double bonds belonging to the center part of the molecules. The features at 1439 and 1395~\cm\ can be referred to the wiggling and torsion of the CH$_3$ group. The low-lying resonances at 1269 and 1243~\cm\ belong to an asymmetric stretch oscillation of the C-O-C bond. The position of the detected features perfectly coincides with the resonance frequencies determined by Huang and Shih \cite{Huang06}.

\begin{figure}
	\centering
		\includegraphics[width=1\textwidth]{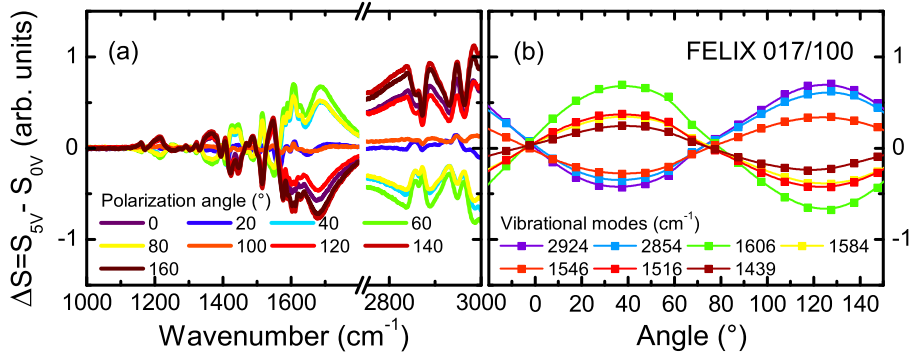}
	\caption{(a)~Spectra of the intensity difference $\Delta S= S_{5V}-S_{0V}$ for various polarization states of the incident light measured at room temperature. (b)~Polarization-dependent measurement of the change of the transmitted intensity after creating 5~V on the two windows for seven different vibrational modes. Only the strongest modes are displayed. The maximal change of the transmission appears at $35^{\circ}$, followed by a minimum at $125^{\circ}$. The sinusoidal modification evidences that the liquid crystals react on the applied electric field. }
	\label{fig:5-Angles}	
\end{figure}

Fig.~\ref{fig:5-Angles} shows a change of the infrared spectra when an electric field of $E=5~$~V is applied. All measurements were performed at $25^{\circ}$C. Due to the rearrangement of the liquid-crystal molecules  a strong modification of the intensity could be observed. We can can deduce two different-oriented transition dipole moments: the first belongs to the vibrational mode of the rigid body of the molecule and the other points out from the molecule axis related to the end groups.
The modes containing the CH$_2$ and CH$_3$ bonds exhibit a minimum at $35^{\circ}$ and a maximum at $125^{\circ}$, correspondingly. The minimum of the modes belonging to the molecule body are slightly shifted and can be found at $125^{\circ}$. The green curve, for instance, belongs to a vibrational mode with an aligned dipole moment parallel to the central molecular frame. At $35^{\circ}$ it reaches a maximum for a voltage of 5~V, this means that the dipole moment rotates away from the electric field vector of the infrared light. Hence, the transmitted intensity increases at this specific wavelength. In contrast, the intensity of the dipole moments related to the end groups of the molecule is reduced because the transition dipole moments are aligned parallel to the incident radiation.

\subsection{Time-Dependent Optical Behavior}
\begin{figure}
	\centering
		\includegraphics[width=1\textwidth]{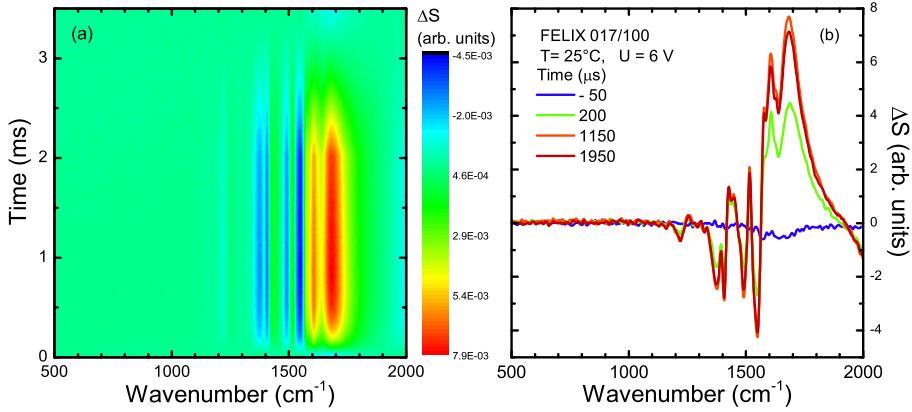}
	\caption{(a)~Contour plot of time-dependent infrared spectra of FELIX, demonstrating the
switching process of the liquid-crystal cell for an electric field pulse of the strength $E=12$~kV/cm with a pulse width of 2~ms. In a spectral region from 1200 to 1800~\cm\ several features can be recognized corresponding to the orientation of their transition dipole moment.
(b)~Time-evolution of the spectra demonstrate for four selected times after the voltage pulse of 6~V was applied. The spectrum (blue), recorded 50~$\mu$s before the pulse is applied, exhibits almost no change and is flat. Several features appear at time 200~$\mu$s that can be ascribed to certain vibrational modes. The signal saturates after 1150~$\mu$s and shows no further variation of the intensity.}
	\label{fig:6-Contour}	
\end{figure}

In order to investigate the dynamics of the liquid-crystal switching process the polarizer was fixed at an angle of $45^{\circ}$, and voltages between 4 and 6~V applied between the two plates. The time-dependent infrared signal is presented in a contour plot in Fig.~\ref{fig:6-Contour}. The rectangular voltage pulse had a width of 2~ms with a repetition rate of 30~Hz and the signal was recorded for a period of 4~ms. To improve the signal-to-noise ratio, five spectra were averaged. The time resolution was set to 25~$\mu$s and the spectral resolution was 2~\cm. We can conclude that the liquid crystal molecules rotate and reorient like a stiff body under the influence of an external electric field.

The extreme values of $\Delta S(\nu,t)$ are displayed in blue or in red in the contour plot. The transition dipole moments $\mu_e$ are aligned parallel to $E_{\rm IR}$ lead to a reduced and, hence to a negative signal whereas in contrast for a positive $\mu_e$ is turned away from $E_{\rm IR}$. This becomes more obvious in Fig.~\ref{fig:6-Contour}(b) where spectra at different times are plotted. As a check, data were taken $50~\mu$s before the pulse; no variation of the vibrational modes and no shift is observed. After applying the voltage pulse, however, the spectrum dramatically alters and reveals the feature exactly as in Fig.~\ref{fig:5-Angles}(a) which marks the final state. Also the absolute values of the change are the same. In accord with Fig.~\ref{fig:4-Transmission}(a) and \ref{fig:5-Angles}(a) no variation can be identified below 1000~\cm\ because the ITO layer absorbs most of the incident light. The switching process is completed after about 1~ms and the signal stays constant over a period of 800~$\mu$s.

\begin{figure}
	\centering
		\includegraphics[width=1\textwidth]{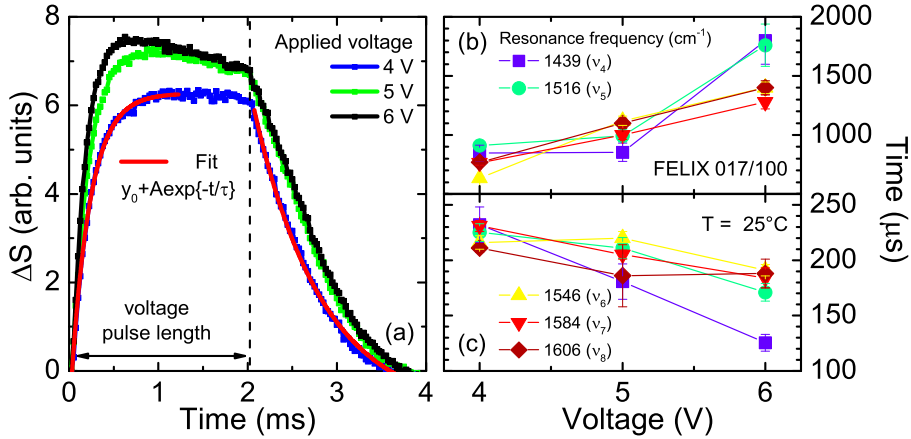}
	\caption{(a)~The variation of the intensity $\Delta S(\nu,t)$ is displayed for the resonance frequency 1606~\cm\ for three different voltages. The rise of the signal as well as the drop can be fitted well  by a single exponential function $y_0+A\exp(-t/\tau)$. The signal increases with increasing voltage. At the same time the switching speed increases. In contrast to that the relaxation process retards with higher  voltage. (b)~Voltage-dependence of the rise time $\tau_{\rm or}$ for five different resonance frequencies. With increasing voltage the switching time decreases continuously.
(c)~Whereas the relaxation time $\tau_{\rm reor}$ deceases, but no direct dependence on the voltage can be discovered. For 6~V $\tau_{\rm reor}$ reveals a larger variance.}
	\label{fig:7-Time}	
\end{figure}
To determine the rise and fall time, the temporal evolution of $\Delta S(\nu, t)$ of five vibrational modes were used. The change of the intensity of the $\nu_8$ mode is plotted as a function of time in Fig.~\ref{fig:7-Time}(a) for three different voltages. With a rise time of less than 10~ns,
the voltage pulse basically has a rectangular shape; however, the infrared signal reacts slowly and requires several hundreds of microsecond until it saturates. Due to the viscosity of the liquid crystals the molecules react with a certain delay to the electric field. The switching time $\tau_{\rm or}$ diminishes with rising voltage whereas the amplitude increases slightly. However, the signal overshoots at the beginning for 6~V and relapses back to the value of the 5~V pulse. The rise time can be determined by fitting the experimental data with a single exponential function $y_0+A\cdot\exp\left\{-t/\tau\right\}$. The extracted values of $\tau_{\rm or}$ are summarized in Fig.~\ref{fig:7-Time}(b) and (c) as a function of the applied voltage.
At the end of the pulse the signal relaxes slowly back to its initial values within 2~ms and is at least by a factor of four larger than the rise time $\tau_{\rm or}$. The time for reorientation $\tau_{\rm reor}$ can be determined from the temporal evolution of the recovery process by a single exponential function.

At a voltage of 4~V almost all vibrational modes reveal a switching time $\tau_{\rm or}$ between 210~$\mu$s and 230~$\mu$s; with increasing voltage this range enlarges to maximum 70~$\mu$s for 6~V whereas the values  are between 190~$\mu$s and 120~$\mu$s. The various time constants imply that constituents of the molecules  react differently on the electric field. The vibrational modes related to the molecule body ($\nu_8$-mode) rotate slower than the ones ($\nu_4$-mode) which are connected to the end groups. The theoretically calculated value for $\tau_{\rm or}= 170~\mu$s (for 5~V) derived from Eq.~(\ref{eq:tau}) agrees very well with the experimentally determined $\tau_{\rm or}$ which is of about 200~$\mu$s. The reason for the small discrepancy may be due to uncertainties in the distance of the spacer plates or in the material specific parameters $\eta$ and $\vec{p}_m$; they all depend strongly on temperature. From our limited data set, we cannot verify the predicted $1/E$ characteristic of the rise time; our data indicate a constant time. Probably,  $\tau_{\rm or}$ exhibits the expected field dependence and the divergent behavior for smaller voltages.

As it was expected from the theoretical calculations of $\tau_{\rm reor}$
the relaxation times in Fig.~\ref{fig:7-Time}(c) are larger than the switching time. A time constant of 40~ms was predicted but not overserved. Furthermore, the relaxation rate is a function of the applied voltage but according to Eq.~(\ref{eq:tau}) it should be independent of any external parameter. One possible explanation therefor is the stronger tilting of the molecules with increasing voltage as well as the total amount of realigned molecules increases as well. Thereby, the layer of oriented molecules increases which leads to a kind of self-stabilization effect resulting in the extended decay. For this reason, the relaxation rate $\tau_{\rm reor}$ is not a function of the intrinsic elastic constant $K$, but also of the external eclectic field.

\section{Optical Switching}
\label{sec:OpticalSwitching}
Over the last decades photo-induced phase transitions were investigated in several material classes, for instance, polymers \cite{Koshihara90a},  organic charge-transfer salts \cite{Koshihara90b,Okamoto04,Iwai06a,Ikegami07,Okamoto06,Mitrano14}, transition-metal oxides like vanadium oxides \cite{Cavalleri04}, cuprates \cite{Demsar99}, and manganites \cite{Rini07}.
In the following we study the one-dimensional organic mixed-stack crystals tetrathiavulvalene-chloranil (TTF-CA) that undergoes a ionic-neutral phase transition at $T_{\rm NI}=81.5$~K that can also be induced by light. Initiated by the groundbreaking experiments by Koshihara {\it et al.} \cite{Koshihara90a},
ultrafast pump-probe experiments have been performed to examine the photo-induced phase transition {PIPT} in the femto- and picoseconds time range \cite{Koshihara90a,Koshihara99,Suzuki99,Iwai06a}.
The understanding of these phenomena was boosted by theories of Nasu and Yonemitsu \cite{Nasu97,Nasu04,Yonemitsu06,Yonemitsu08}.
In the case of TTF-CA open questions concern the relaxation of metastable domains, the related time scale, and the modification of the infrared spectrum due to photo-excitation;
some of these have recently be addressed by time-resolved FTIR-spectroscopy \cite{Peterseim15}.

\subsection{Neutral-Ionic Transition in TTF-CA}
\label{sec:TTF-CA}
\begin{figure}[h]
	\centering
		\includegraphics[width=1\textwidth]{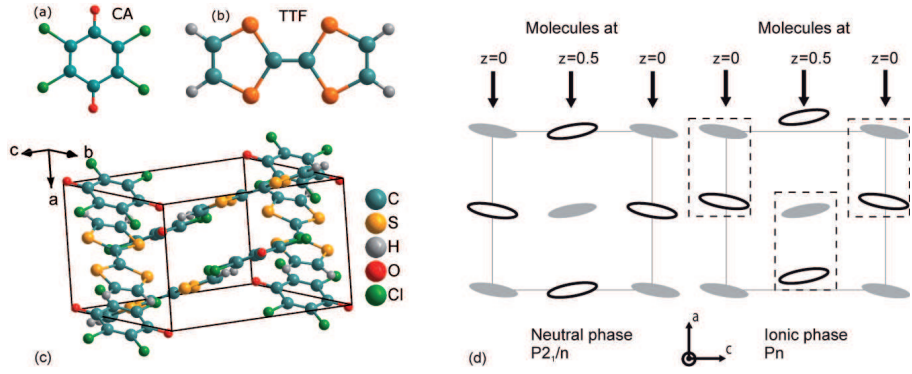}
	\caption{(a)~Chloranil molecule (CA, C$_6$Cl$_4$O$_2$) and (b)~tetrathiafulvalene (TTF, C$_6$S$_4$H$_4$). (c)~Monoclinic unit cell of TTF-CA at room temperature. The TTF and CA molecules are ordered along the crystallographic $a$-axis. (d)~At $T=300$~K the space group of the unit cell is P2$_1$/n and the CA and TTF molecules are stacked equally spaced along the $a$-axis. A further stack is located at $z=c=0.5$, respectively, at which the TTF-CA pairs are tilted opposite to the $a$-axis.
In the ionic phase the TTF and CA molecules dimerize along the stack. By the charge transfer of about $\rho=0.6~e$ electric dipoles are formed along the stacking direction. }
	\label{fig:10-TTF-CA}	
\end{figure}
At ambient conditions the planar molecules of TTF and CA are equidistantly arranged in alternating stacks along the $a$-direction (Fig.~\ref{fig:10-TTF-CA}). Since the charge transfer  $\rho=0.2~e$ as determined from optical experiments is rather small \cite{Girlando82,Girlando83,Jacobsen83}, this state is referred to as the neutral phase; upon cooling the charge transfer increases slightly. At $T_{\rm NI}=81.5$~K a phase transition occurs where the molecules dimerize along the stack with intermolecular distances 3.504~{\AA} and 3.685~{\AA} \cite{LeCointe95}, and the ionicity $\rho$ increases from 0.3~$e$ to about 0.6~$e$; Fig.~\ref{fig:10-TTF-CA}(d) displays a sketch of the arrangement.
Accordingly the dielectric constant diverges at the transition with a pronounced frequency dependent response \cite{Tomic15}. The application of 11~kbar  pressure shifts the transition to room temperature \cite{Mitani87,Dengl14}.

\begin{figure}
	\centering
		\includegraphics[width=0.8\textwidth]{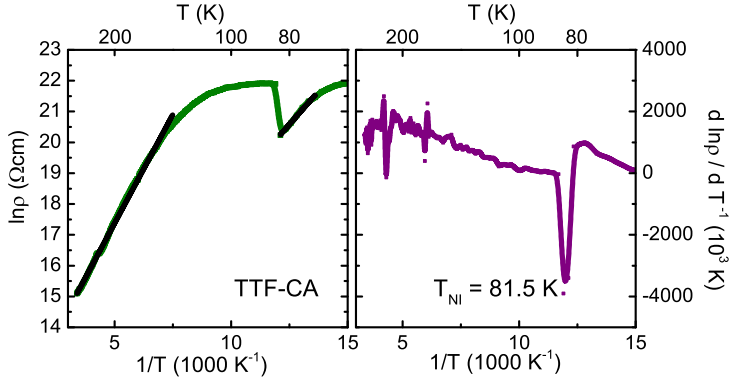}
	\caption{Arrhenius plot of the resistivity  $\rho_{\rm dc}(T)$ (green) of TTF-CA along the stacking direction. Above the transition it behaves as a classic band insulator with an activation energy of $\Delta=0.12$~eV.}
	\label{fig:11-TTF-CA-rho}	
\end{figure}
The resistivity of TTF-CA along the stacking direction increases upon cooling,
following an activated behavior as displayed in the Arrhenius plot of Fig.~\ref{fig:11-TTF-CA-rho}.
At $T_{\rm NI}$, $\rho_{\rm dc}(T)$ drops by one order of magnitude before it increase again at lower temperatures. The strong enhancement of the conductivity is attributed to the rising number of neutral-ionic domain walls in the neutral phase which also explains the dielectric behavior and non-linear transport \cite{Okamoto91}. Alternatively it was suggested, that
the multiphonon Peierls coupling leads upon approaching the phase transition to a progressive shift of spectral weight and of the coupling strength toward the phonons at lower frequencies, ending in a soft-mode behavior only for the lowest-frequency phonon near the transition temperature \cite{Masino06,Girlando08}.  In the proximity of the phase transition, the lowest-frequency phonon becomes overdamped due to anharmonicity induced by its coupling to electrons.

\begin{figure}
	\centering
		\includegraphics[width=1\textwidth]{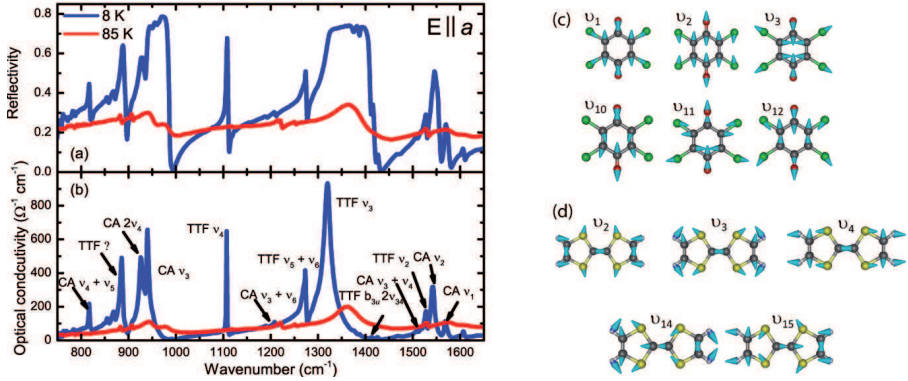}
	\caption{(a)~Reflectivity and (b)~optical conductivity of TTF-CA for $T=8$~K (blue) and 85~K (red) taken for the polarization along the stack. The maximum is mainly caused by emv-coupled modes which gain intensity in the dimerized ionic phase. (c) Moleuclar vibrations of chloranil (CA) and (b) tetrathiofulvalene (TTF):  the upper rows depict the gerade a$_g$-modes and the lower rows the ungerade b$_{1u}$-modes.}
\label{fig:13-TTF-CA-refmodes}	
\end{figure}

%\subsection{Steady-State Optical Properties}
In Fig.~\ref{fig:13-TTF-CA-refmodes} the mid-infrared reflectivity and the optical conductivity of TTF-CA are presented for temperatures above and below the neutral ionic transition. In this energy range the intramolecular modes of the TTF and CA molecules can be identified and assigned \cite{Girlando78,Girlando83,Bozio79,Peterseim15}; some of them are sketched in panels (c) and (d).
Along the $a$-direction the symmetric a$_g$ as well as the infrared-active b$_{3u}$ modes of CA and TTF can be observed whereas the first one is only infrared active due to the emv-coupling. Above the phase transition in the neutral phase the TTF and CA molecules are not dimerized and thus the a$_{g}$ modes are only weakly infrared-active and the optical conductivity is low.

In the ionic phase the point inversion symmetry is lost and the molecules become dimerized: the intensity of a$_g$ modes is enhanced enormously. Time-resolved measurements thus allow us to explore the dynamics at the neutral-ionic phase transition by looking at the change in dimerization of the TTF and CA molecules, the increase of the ionicity. We can investigate the building of metastable domains until they eventually annihilate.

\subsection{Experimental Details}
For photo-induced experiments a Nd:YAG laser (B.M Industries/Thales, YAG\--502\-DNS\--DPS920)
was operated in the second harmonics $\lambda=532$~nm (2.35~eV) of the fundamental wavelength (1064~nm).
The pulse length is 8~ns. The repetition rate can be selected internally and externally between 1~Hz and 20~Hz. The laser intensity is adjusted continuously by a Brewster plate in addition to  neutral density filters between 0.1 and 0.5 optical density. The laser intensity was checked by a power and energy meter in front of the sample. The long term stability of the laser power is better than 5\%.

\begin{figure}
	\centering
		\includegraphics[width=0.8\textwidth]{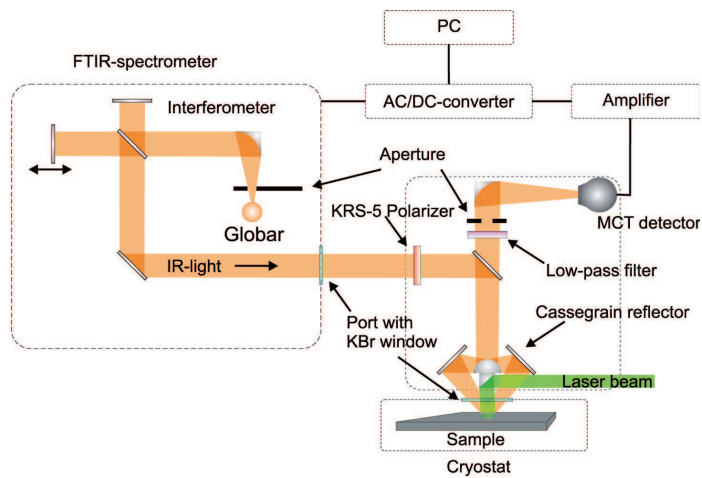}
	\caption{Sketch of the optical setup including the laser beam to excite the sample. The setup consists of three main parts (dashed lines). The centerpiece of the setup is the spectrometer containing the electronics as the A/D-converter and amplifier, but also the Globar light source, the interferometer with the beamsplitter and several mirrors to deflect the light beam. The second part is the infrared microscope which is attached to the spectrometer. There, the light is focused on the sample by a Cassegrain reflector. Furthermore, a polarizer and bandpass filter can be mounted in the microscope.  The MCT detector is placed at the end of the light beam. }
	\label{fig:7-Setup}	
\end{figure}

\begin{figure}
	\centering
		\includegraphics[width=0.57\textwidth]{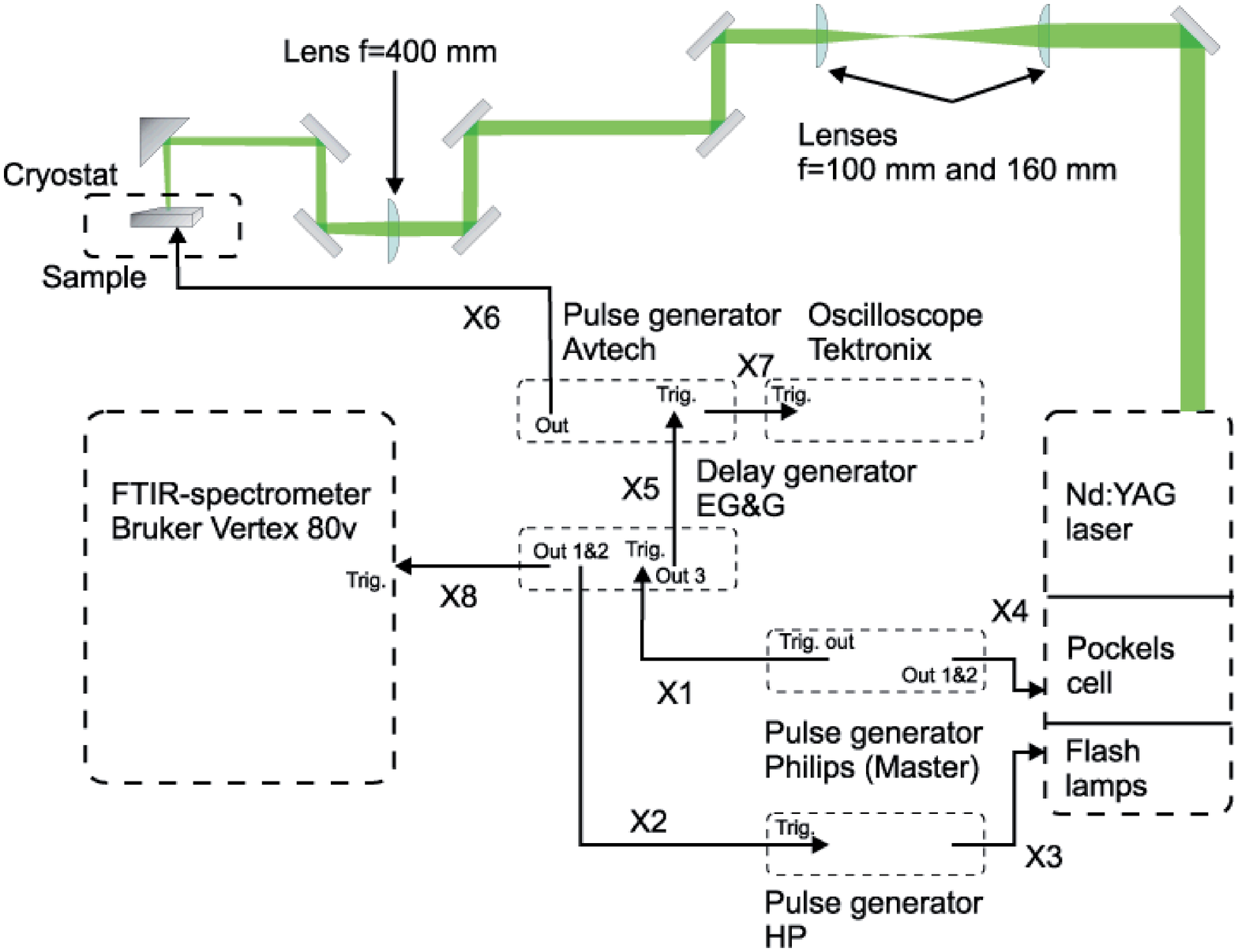}\hspace*{5mm}
		\includegraphics[width=0.36\textwidth]{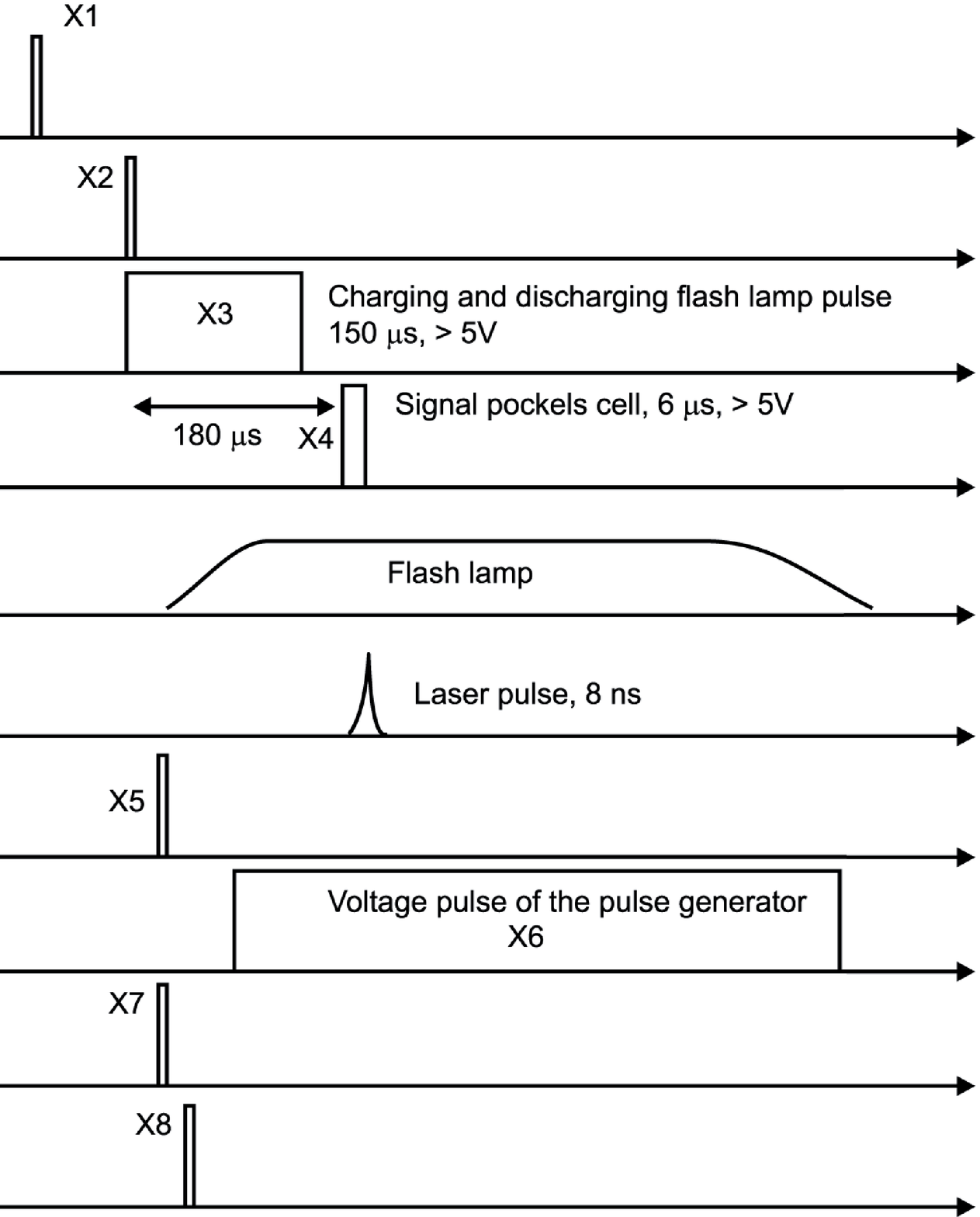}
	\caption{The left panel sketches the laser setup. The centerpiece is the pulsed Nd:YAG laser which is externally controlled by two synchronized pulse generators. A delay generator is additionally implemented in the setup to guarantee synchronization between the laser pulse and the voltage pulse generator as well as the oscilloscope. It also triggers the spectrometer to acquire the infrared data. The laser beam (green) is collimated, directed and focused on the sample via several mirrors and lenses.
%X1 ... X8 label the signal which are used to trigger the various instruments.
The pulse sequence to control the laser-induced experiment is depicted on the right panel. X1 is the master trigger pulse triggering the delay generator. It synchronizes the flash lamps (X2 and X3), the voltage pulse generator (Avtech) (X5) and the  FTIR-spectrometer (X8) with each other. X4 controls the Pockels cell in the resonator with a minimum length of 6~$\mu$s. The pulse X6 is the voltage pulse which is applied to the sample for the photoconductivity measurements, for instance. The oscilloscope records the variation of the sample current or resistance and is synchronized with the other instrument via the signal X7.   }
	\label{fig:8-Laser}	
\end{figure}

A telescope arrangement reduces the diameter of the laser beam from 1~cm by a factor of 2, illustrated in Fig.~\ref{fig:8-Laser}. The laser beam is directed from the optical bench to the infrared microscope via several mirrors (see Fig.~\ref{fig:7-Setup}). There, it is deflected on the sample by a $45^{\circ}$ aluminum coated mirror mounted below the Schmidt-Cassegrain objective of the Bruker HYPERION infrared microscope. A lens ($f=400$~mm) focuses the beam on the sample, by varying the focal length. The light is circular polarized. The laser system and optical setup are spatially decoupled from the spectrometer and mounted on an optical table to suppress possible external vibrations.
		
The laser pulse sequence is controlled externally to ensure the temporal synchronization between the laser, the pulse generators and the FTIR-spec\-tro\-meter, as depicted in Fig.~\ref{fig:8-Laser}. The charging of the flash lamps is triggered by an external signal as well as the Pockels cell generating the laser pulse. Therefore, a HP pulse generator (PM 5786 B) sends the trigger signal X1 to the delay generator (EG\&G Princeton Applied Research Model 4144). One of the delay generator output signals X2 is forwarded to a second pulse generator (HP 214B) which releases a further delayed pulse X3 with a minimum length of 150~$\mu$s and minimum height of 5~V. It initializes the charging of the flash lamps of the laser system. After signal X3 drops to zero the lamps are charged with a delay of 1.5~ms. It also activates the discharge of the lamp with a delay of 15~$\mu$s. About 30~$\mu$s after the end of the charging and discharge pulse X3 a further voltage signal X4 from the first HP PM 5768 B pulse generator (length 6~$\mu$s, 6~V) is sent to the Pockels cell generating the laser pulse. In the case of the photoconductivity measurements the delayed trigger signal X5 from the delay generator activates the pulse generator (Avtech Electrosystems Ltd., Ottawa) to apply a voltage pulse X6 to the sample. The synchronizing  pulse X7 of the Avtech device goes to the Tektronix oscilloscope to start the acquisition of the photocurrent. The third signal X8 from the delay generator is used to initialize the time-resolved infrared measurement of the FTIR-spectrometer.

\subsection{Time-Resolved Infrared Studies}
\label{sec:TTF-CA_PIPT}
For the PIPT measurements we excite only one of the first transitions of TTF$^+$ by the second harmonic of the Nd:YAG laser; all intermolecular transitions of CA$^-$ lie above present photon energy. Also excitations from lower lying bands into the valence band are possible. Suzuki {\it et al.} compared the dependence of the conversion efficiency on the photon energy; by excitation of intramolecular transitions no threshold intensity occurs to create neutral domains \cite{Suzuki99}.

\begin{figure}
	\centering
		\includegraphics[width=.5\textwidth]{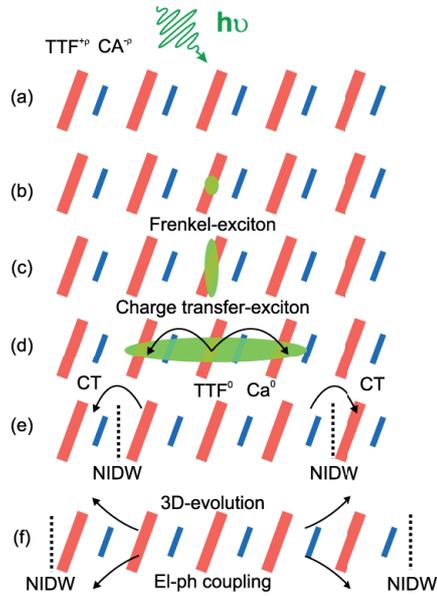}
	\caption{(a)~Illustration of a one-dimensional chain of dimerized TTF$^{+\rho}$ and CA$^{+\rho}$ pairs in the ionic phase in TTF-CA. A TTF$^{+\rho}$ molecules is excited with a laser pulse with the photon energy $h\nu$. (b)~Vertical excitation of the HOMOs of TTF$^{+\rho}$ according to the Frank-Condon-principle in the LUMO of TTF$^{+\rho}$. The excitation is strongly localized to the molecule. (c)~Creation of excitons, for instance Frenkel-excitons, which are delocalized across the whole molecules. (d)~Via different relaxation processes and channels charge transfer excitons are created which triggers to the transition of the neutral phase.
(e)~The dimerization is suppressed and the charge between the molecules is redistributed. A neutral domain is created in the ionic host matrix which is separated by neutral-ionic domain walls.
(f)~Afterwards the neutral domains extend along the one-dimensional chain. By electron-phonon-coupling also neighboring ionic chains are converted into neutral regions and a three-dimensional domain is established. }
	\label{fig:14-PIPT}	
\end{figure}
In Fig.~\ref{fig:14-PIPT} the optical generation of neutral domains is schematically illustrated. The light pulse will create intramolecular Frenckel-excitons, which are electron-hole pairs strongly localized on the excited molecules. They  decay via various channels into several charge-transfer excitons \cite{Nasu04} and by that lead to a phase transition. In the case of direct excitation only one exciton is created, which is not enough to establish a macroscopic, metastable domain extended over several D$^0$A$^0$ pairs. With a sufficient number of photons a multiplicative, non-linear effect can be established, formating metastable domains. By the creation of the one-dimensional, neutral, non-dimerized region, neutral-ionic domain walls are formed between the neutral and ionic parts. The excitation energy of these domain walls  is about 0.1~eV \cite{Mitani87,Okamoto91} and corresponds to the activation energy of 120~meV and 65~meV in the ionic phase \cite{Peterseim15}.
This is in good agreement with predictions of an activation energy between 25 and 56~meV \cite{Nagaosa86a,Nagaosa86b,Soos07}.
Afterwards, the electronic system couples to the crystal lattice and exciting phonons \cite{Iwai06a} via electron-phonon coupling and hence, creates shock-waves. They can convert neighboring chains into neutral domains.

\begin{figure}
	\centering
		\includegraphics[width=1\textwidth]{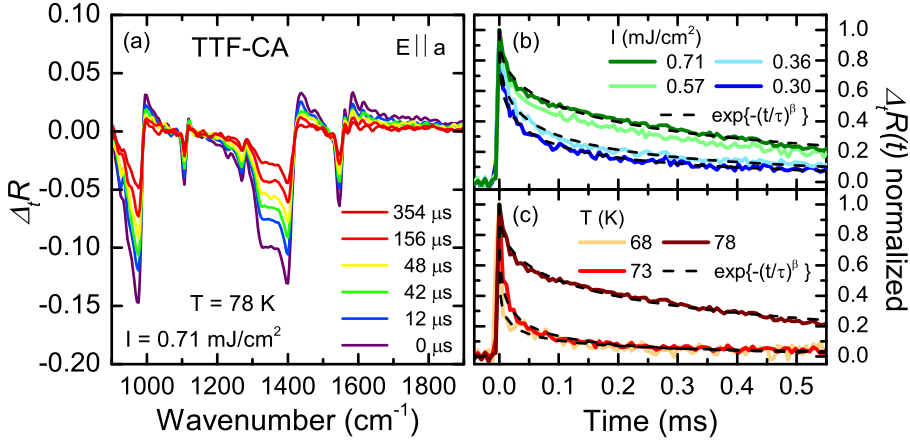}
	\caption{(a)~The temporal sequence of the reflectivity change $\Delta_t R=R(t)-R(0)$ (solid lines) after photo excitation is depicted for various times. The behavior resembles the static reflectivity difference $\Delta_T R =R_{\rm 85 K} -R_{\rm 79 K}$ presented in Fig.~\ref{fig:13-TTF-CA-refmodes}(a).
(b)~and (c)~Normalized $\Delta_t R(t)$ for  various laser intensities recorded at 1390~\cm\ for $T=78$~K and for different temperatures for 0.71~mJ/cm$^2$. The time profile can be successfully fitted by a stretched explonential function $\exp\{-({t}/{\tau})^{\beta}\}$ (dashed linies).}
	\label{fig:14-TTF-CA-transient}	
\end{figure}
In Fig.~\ref{fig:14-TTF-CA-transient}(a) the time-dependent behavior of the reflectivity change  $\Delta_t R=R(t)-R(0)$ at $T=78$~K is plotted for a laser pulse intensity of 0.71~mJ/cm$^2$.
It directly compares to the static reflectivity change $\Delta_T R =R_{\rm 85 K}-R_{\rm 79 K}$ plotted in Fig.~\ref{fig:13-TTF-CA-refmodes}(a). By photo excitation $\Delta_t R$ becomes negative within a short time which is below the experimental time resolution of 6~$\mu$s. The direct comparison of the $\Delta_t R$ shape and the static reflectivity change $\Delta_T R$ reveal that the ionic phase was not only dissolved, but also a transition into a neutral state was induced. Within several hundreds of microseconds the signal $\Delta_t R$ relaxes back to zero which means that the ionic phase is reestablished. Moreover, no change of the spectral shape with the elapsed time and laser pulse intensity could be detected.
By comparing the shape of $\Delta_t R$ with the corresponding difference of the reflectivity $\Delta_T R$ between 78~K and 85~K, we conclude that the vanishing of this features indicates the dissolving of the dimerization state between the TTF and CA molecules. Moreover, we suggest that metastable, non-dimerized neutral domains in the ionic matrix are created.

To trace the temporal evolution of the PIPT in dependence of the pump intensity and the sample temperature, we have chosen the very intense $\nu_3$ (a$_{\rm g}$) mode of TTF residing at 1390~\cm\ since we have asserted that the temporal evolution is the same for the whole spectra. The normalized $\Delta_t R(t)$ is represented in Fig.~\ref{fig:14-PIPT}(b) for different pulse intensities.
At the beginning the signal decays very fast and at the end it flattens out. At the vicinity of $T_{\rm NI}$ the first component decays faster with decreasing laser intensity.
A fit by a simple single or double exponential function \cite{Nasu04} does not yield satisfactory results. However, we obtain an excellet agreement when using a stretched-exponential function, which is also called Kohlrausch-William-Watt function
\begin{equation}
\Delta_t R(t) \propto  \exp\left\{-({t}/{\tau})^{\beta}\right\}
\label{eq:KWW} \quad ,
\end{equation}
as depicted in Fig.~\ref{fig:14-TTF-CA-transient}(b) and (c).
The fitting parameters $\beta$ and $\tau$ are a function of the laser intensity and decreases from 0.35 to 0.42 and from $3.4\times 10^{-5}$ to $2.4\times 10^{-4}$~s with decreasing laser intensities.

In Fig.~\ref{fig:14-TTF-CA-transient}(c) $\Delta_t R(t)$ is displayed for various temperatures $T=68$, 73, and 78~K. Far below $T_{\rm NI}$, the temporal dynamics of the reflectivity drops very fast within the first 20~$\mu$s and approaches asymptotically a constant value which is in contrast to the temporal profile at the vicinity ($T$=78~K) of $T_{\rm NI}$ which constantly diminishes. Similar to the dependence of the fitting parameters on the laser intensity the effective recombination time $\tau$ as well as the stretching exponent decrease from $2.4\times 10^{-4}$~s to $3.2 \times 10^{-6}$~s and from 0.42 to 0.23, respectively, with decreasing sample temperature.

The observed time profile can be explained by a random-walk annihilation process of the
generated neutral-ionic domain walls. Our comprehensive
time-resolved infrared study and the random-walk
model \cite{Peterseim15} allow us to conclude that close to the phase transition, large domains are
formed due to the valence instability.
We find that the merger and interaction of
the induced domains play an important role for the formation
of the macroscopic domains and deduce from the
model with decreasing laser intensity, the average domain
size decreases. At lower temperatures the ionic phase is
more robust; the average domain size is much smaller and
changes less with laser intensity. The random walk of the neutral-ionic domain
walls is the dominant factor for the relaxation of
the metastable domains in the temperature range considered.

\section{Summary}
We have presented two examples where time-resolved infrared investigations using step-scan Fourier-transform spectroscopy provide insight in the molecular dynamics at the phase transition and molecular orientation.
In the case of liquid crystals, the application of an electric field switches the orientation of the molecular dipoles changing the transmission of the cell for polarized infrared light. Watching the time evolution of the pronounced molecular vibrational modes allows us to extract the time constant of approximately 2~ms and its dependence on the applied voltage.
The neutral-ionic phase transition of TTF-CA can be photo-excited with a short laser pulse, creating domain walls that are mobile and eventually annihilate. We follow the time-dependence of the vibrational modes activated by the changed dimerization and ionicity. The relaxation back to the initial state strongly depends on the temperature and laser intensity, it extends from a 3~$\mu$s to almost 1~ms.
For both examples we give details of the experimental setups and limitations. We demonstrate the applicability of step-scan Fourier-transform spectroscopy in a large dynamical range.

\begin{acknowledgements}
We thank F. Sch{\"o}rg and F. Giesselmann for providing the liquid crystals and G. Untereiner for continuous help; E. Kurz and N. Fr{\"u}hauf supported us when building the cell and use of their clean room. Funding by the Deutsche Forschungsgemeinschaft (DFG) is acknowledged. T.P. thanks the Carl-Zeiss-Stiftung for financial support.
\end{acknowledgements}

%\bibliography{Pulse}   % name your BibTeX data base

%\begin{thebibliography}{}

% etc
%\end{thebibliography}

\end{document}